\documentclass[twocolumn,aps,prl,superscriptaddress,floatfix, nofootinbib]{revtex4-1}
\usepackage{graphicx}
\usepackage{amsmath}
\usepackage{natbib}
\usepackage{gensymb}
\usepackage{afterpage}
\usepackage{braket}

\begin{document}

\title{Strong Coupling between Microwave Photons and Nanomagnet Magnons}
\author{Justin T. Hou}
\affiliation{Department of Electrical Engineering and Computer Science, Massachusetts Institute of Technology, Cambridge, MA 02139, USA}
\author{Luqiao Liu}
\affiliation{Department of Electrical Engineering and Computer Science, Massachusetts Institute of Technology, Cambridge, MA 02139, USA}

\date{{\small \today}}

\begin{abstract}

Coupled microwave photon-magnon hybrid systems offer promising applications by harnessing various magnon physics. At present, in order to realize high coupling strength between the two subsystems, bulky ferromagnets with large spin numbers are utilized, which limits their potential applications for scalable quantum information processing. By enhancing single spin coupling strength using lithographically defined superconducting resonators, we report high cooperativities between a resonator mode and a Kittel mode in nanometer thick Permalloy wires. The on-chip, lithographically scalable, and superconducting quantum circuit compatible design provides a direct route towards realizing hybrid quantum systems with nanomagnets, whose coupling strength can be precisely engineered and dynamic properties can be controlled by various mechanisms derived from spintronic studies.

\end{abstract}
\maketitle
Hybrid quantum systems have been extensively studied to harness advantages of distinct physical systems and realize functions that cannot be achieved with any individual sub-system alone \cite{xiang_hybrid_2013, kurizki_quantum_2015}. In particular, cavity and circuit quantum electrodynamics (Cavity/Circuit QED) \cite{wallraff_strong_2004, schoelkopf_wiring_2008, wendin_quantum_2017} provide promising platforms for realizing hybrid quantum systems using Josephson qubits, mechanical systems \cite{aspelmeyer_cavity_2014}, atoms \cite{buluta_natural_2011}, quantum dots \cite{petersson_circuit_2012}, as well as ensembles of spins \cite{imamoglu_cavity_2009, schuster_high-cooperativity_2010}. For realizing coherent energy and information exchange, electric dipole interactions have been traditionally utilized to couple photons with other quantum excitations. Recently, coupled microwave photon-magnon systems have received great attention as an alternative approach to realize strong light-matter interactions using magnetic dipole coupling \cite{soykal_strong_2010, huebl_high_2013, zhang_strongly_2014, tabuchi_hybridizing_2014, goryachev_high-cooperativity_2014, bai_spin_2015,  morris_strong_2017}. In this system, magnons in magnetic materials with high spin density are utilized, where coupling strength $g$ is collectively enhanced by square root of number of spins ($g=g_s\sqrt{N}$) \cite{tabuchi_hybridizing_2014} to overcome the weak coupling strength $g_s$ between individual spins and the microwave field. Along these lines, sizable ferrimagnets, yttrium iron garnet (YIG) with millimeter dimensions have been employed for reaching strong coupling. While great success has been demonstrated in achieving coherent sensing and control over the magnonic quantum state using this architecture \cite{tabuchi_coherent_2015,zhang_magnon_2015, lachance-quirion_resolving_2017}, one important question remains unanswered: whether such a system is scalable for achieving integrated hybrid quantum systems. In the meantime, reducing the size of magnets in this hybrid quantum system can potentially provide another degree of freedom for realizing active sensing and control of quantum states. In the study of spin electronics, sophisticated techniques have been developed for manipulating and detecting spin states using various electrical methods; however, these effects work efficiently only in nanoscale magnets \cite{kiselev_microwave_2003, ando_electric_2008, demidov_magnetic_2012, liu_magnetic_2012, duan_nanowire_2014}. In this Letter, by utilizing lithographically defined superconducting resonators, we demonstrate strong magnon-photon coupling with a nanometer size Permalloy thin film stripe (Permalloy=Py=$\text{NiFe}$), where the number of spins is on the order of $10^{13}$, 3 orders of magnitude lower than previous studies. The realization of magnon-photon coupled systems using metallic ferromagnets with conventional Si-substrates demonstrates a highly engineerable and industrial compatible on-chip device design. Moreover, the large coupling strength with nanomagnets provides a direct avenue towards scalable hybrid quantum systems which can benefit from various magnon physics, including nonlinearity \cite{wang_bistability_2018}, synchronized coupling \cite{harder_level_2018, zare_rameshti_indirect_2018}, non-Hermitian physics \cite{zhang_observation_2017, harder_topological_2017}, as well as current or voltage controlled magnetic dynamics \cite{kiselev_microwave_2003, ando_electric_2008, demidov_magnetic_2012, liu_magnetic_2012, duan_nanowire_2014}.

\begin{figure*}[!tbp]
\includegraphics[width=18cm]{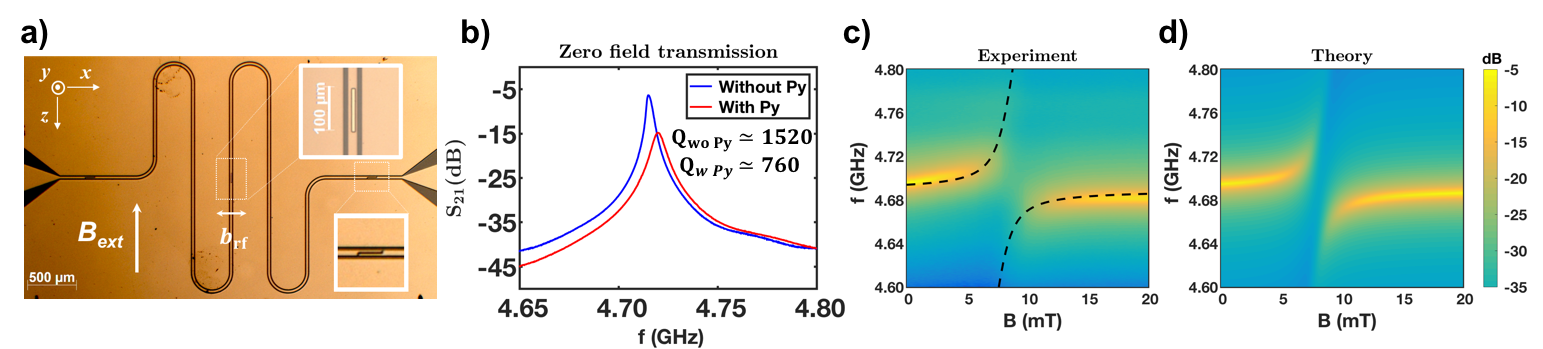}
    \caption{{\bf a)} Photo of a CPW resonator device with 100$\mu$m$\times8\mu$m$\times50$nm MgO/Py/Pt stripe deposited at the center of the signal line. The distance $l=12$ mm between the two open gaps at the ends of CPW defines the fundamental resonant frequency to be 4.96 GHz.  {\bf b)} Microwave transmissions at 1.5 K with zero applied field before and after Py is deposited, exhibiting resonance signals with quality factors 1570 and 760, respectively. With Py, the resonator mode is blue detuned due to residual coupling to the magnon mode. {\bf c)} Microwave transmission as a function of frequency and in-plane magnetic field at 1.5 K for a sample with 500$\mu$m$\times8\mu$m$\times50$nm Py showing characteristic anti-crossing of magnon-photon coupling. (The data from {\bf b)} and {\bf c)} are taken from different samples). {\bf d)} Theoretical microwave transmission spectrum calculated using input-output theory with parameters obtained from the experiment.}
    \label{fig1}
\end{figure*}

Figure \ref{fig1}(a) shows the image of a typical lithographically defined Nb superconducting coplanar waveguide (CPW) resonator that is coupled with a thin film ferromagnet (inset).  We can model this hybrid system quantum mechanically as a macrospin coupled to an LC resonator through oscillating magnetic field $\mathbf{b_{rf}}$ generated by the inductor, where $\mathbf{b_{rf}}=b_{\text{rf}}\mathbf{\hat{x}}$ is the magnetic field experienced by the macrospin per unit inductor current. During the experiment, an external field $\mathbf{B_{ext}}=-B_0 \mathbf{\hat{z}}$ is applied to tune the intrinsic resonant frequency of the macrospin. The total Hamiltonian of the system can be written as \cite{soykal_strong_2010, goryachev_high-cooperativity_2014,SI}:
\begin{equation}
\hat{H}=\hbar \omega_r (\hat{a}^{\dagger}_r\hat{a}_r+\frac{1}{2})-\omega_m(B_0)\hat{S}_z+g_s(\hat{S}_+\hat{a}^{\dagger}_r+\hat{S}_-\hat{a}_r)  \label{totalH}
\end{equation}
where $\hat{a}^{\dagger}_r $ ($\hat{a}_r$) is the creation (annihilation) operator of microwave photon modes in the resonator and $\mathbf{\hat{S}}=\frac{1}{2}(\hat{S}_{+}+\hat{S}_{-}) \mathbf{\hat{x}}+\frac{1}{2i}(\hat{S}_{+}-\hat{S}_{-}) \mathbf{\hat{y}}+\hat{S}_z \mathbf{\hat{z}}$ is the macrospin operator, with $\hat{S}_{+}$ ($\hat{S}_{-}$)  raising (lowering) the $z$-component of the macrospin. The resonant frequencies of the resonator $\omega_r$ and the macrospin $\omega_m(B_0)$ are given by $\omega_r=1/\sqrt{LC}$  and the Kittel formula, separately. The coupling strength between photons and individual spins in the magnetic material $g_s$ can be represented as $g_s=g_{e}\mu_B b_{\text{rf}} \omega_r/\sqrt{8\hbar Z_r}$ \cite{SI,eichler_electron_2017}, with $Z_r=\sqrt{L/C}$ being the characteristic impedance of the LC resonator. The eigenfrequencies of the hybrid system can be calculated as \cite{huebl_high_2013}: 
\begin{equation}
\omega_{\pm}=\omega_r+\Delta/2\pm\sqrt{\Delta^2+4g^2}/2 \label{twomodes}
\end{equation}
where $\Delta=\omega_m(B_0)-\omega_r$ is the detuning and $g=g_s\sqrt{N}$ is the total magnon-photon coupling strength \cite{SI}. Therefore, in order to achieve scalable strong magnon-photon coupling with reduced $N$, it is important to increase the value of $g_s$. For a fixed resonant frequency of the resonator, two strategies can be employed to achieve this: (i) increasing $b_{\text{rf}}$ by adjusting geometry of the inductive wire, or by placing the magnet close to the location with maximum magnetic field in the resonator; (ii) reducing $Z_r$ by utilizing low-impedance resonators with small $L$ and large $C$. Adopting the first strategy in a superconducting CPW resonator, we first realize a relatively high  $g_s^{\text{CPW}}/2\pi=18$ Hz by depositing the Permalloy stripe directly on top of the signal line with a thin insulating insertion, where strong coupling is realized with as few as $10^{13}$ spins. Furthermore, by combining both strategies, we show that very high coupling $g_s^{\text{LE}}/2\pi=263$ Hz can be achieved in a lumped element LC resonator, which allows another two orders of magnitude reduction in spin number $N$ to achieve similar coupling strength.

\begin{figure*}[!tbp]
\includegraphics[width=18cm]{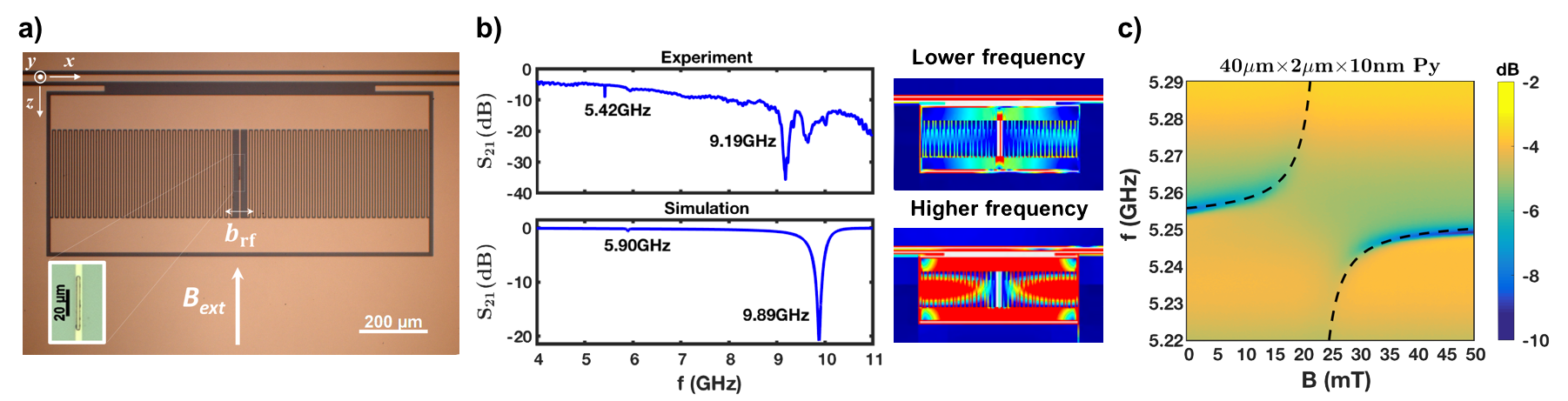}
    \caption{{\bf a)} Image of a low-impedance lumped element resonator device with $40\mu$m$\times2\mu$m$\times10$nm MgO/Py/Ta stripe deposited at the center of the 4 $\mu$m wide inductive wire. {\bf b)} Left: Microwave transmission of the low-impedance resonator obtained from experiment and simulation. Two modes are observed as transmission minima due to photon absorption from the signal line. Right: Simulation of current density distribution of the resononant modes, respectively. Red and blue color indicates regions with strong and weak current densities, respectively. The first harmonic mode exhibits large current density at the central inductive wire which enhances coupling. {\bf c)} Microwave transmission as a function of frequency and in-plane magnetic field at 1.5 K. $\mu_0 M_s=1.1$ T and $g/2\pi=74.5$ MHz are determined from the fitting shown in dashed line. Minimum transmission shows up under resonant conditions.}
    \label{fig2}
\end{figure*}

To enhance the microwave magnetic field generated by unit inductor current $b_{\text{rf}}$, we minimize the width of the CPW resonator signal line in Fig. \ref{fig1}a to be $w^{\text{CPW}}=20$ $\mu$m. The $l=12$ mm long resonator is then capacitively coupled to the external circuit through two gaps at the ends of signal line, leading to a fundamental resonant frequency $\omega_r/2\pi = c/2l\sqrt{\epsilon_{eff}}\approx 4.96$ GHz, where $c$ is the speed of light and $\epsilon_{eff}=6.35$ represents the average dielectric constant of vacuum and Si substrate \cite{goppl_coplanar_2008}. The fundamental mode has a current distribution which reaches maximum at the center of the signal line, where we deposit a MgO(5nm)/Py(50nm)/Pt(10nm) stripe by magnetron sputtering followed by liftoff. The thin insulating MgO layer protects superconducting Nb from ferromagnetic exchange coupling, while bringing Py close to the surface of Nb for large RF magnetic field. We mount the device in a cryostat with a base temperature of 1.5 K and studied the transmission of microwave signal with a Vector Network Analyzer. The transmission of the resonator before and after the deposition of Py stripe at zero field is shown in Fig. \ref{fig1}b, where the quality factor can be determined to be $Q = 1570$ and 760, separately. To tune the frequency of magnetic resonance, an in-plane magnetic field $\mathbf{B_{ext}}$ is applied along the long axis direction of Py stripe. As the RF magnetic field produced by the signal line is perpendicular to the external field direction, the ferromagnetic resonance (FMR) mode can be excited, thereby inducing a microwave photon-magnon coupling. Fig. \ref{fig1}c shows the transmission of a sample with 500$\mu$m$\times$8$\mu$m lateral dimensions as a function of frequency and applied magnetic field. The distinct anti-crossing feature at $B_0=8$ mT is a result of microwave photon-magnon coupling where interaction between the two modes lift the degeneracy in resonance frequencies. The resonant modes evolution can be fitted by Eq.\ \ref{twomodes}, with $\omega_{m}(B_0)=\gamma \sqrt{[B_0+(N_y-N_z)\mu_0 M_s][B_0+(N_x-N_z) \mu_0 M_s]}$ given by Kittel formula, where $\gamma/2\pi \approx 28$ GHz/T is the gyromagnetic ratio. In $\omega_{m}(B_0)$, the demagnetization factors $N_{i}$ are taken into account, which can be analytically calculated with the dimension of the Py stripe \cite{aharoni_demagnetizing_1998}. Through the fitting, we extract the coupling strength $g/2\pi=64$ MHz, the saturation magnetization $\mu_0 M_s=1.2$ T of the Py stripe, and the resonator frequency $\omega_r^{\text{CPW}}/2\pi = 4.690$ GHz. Furthermore, we obtain the decay rates of the resonator mode $\kappa_r/2\pi=1.5$ MHz and magnon mode $\kappa_m/2\pi=122$ MHz by a transmission measurement of bare resonators and an independent FMR measurement of Py(50nm)/Pt(10nm) bilayer, separately. To validate the coupling strength and decay rates, we adopt the input-output theory which gives microwave transmission coefficient as a function of frequency and magnetic field in our system \cite{gardiner_input_1985,huebl_high_2013}:
\begin{equation}
S_{21}(\omega, B_0)=\frac{\kappa_{r,i}}{i(\omega-\omega_r)-\kappa_r+\frac{g^2}{i(\omega-\omega_{m}(B_0))-\kappa_m/2}} \label{ioCPW}
\end{equation}
where $\kappa_{r,i}$ describes the external coupling rate to the cavity and only results in a constant offset in unit of dB. With the parameters measured in our experiment, we plot the theoretical transmission spectrum in Fig. \ref{fig1}d, which attains reasonable agreement with the experiment results, with the minimum transmission signal of the latter limited by the background level.

\begin{figure}[!tbp]
\includegraphics[width=8.5cm]{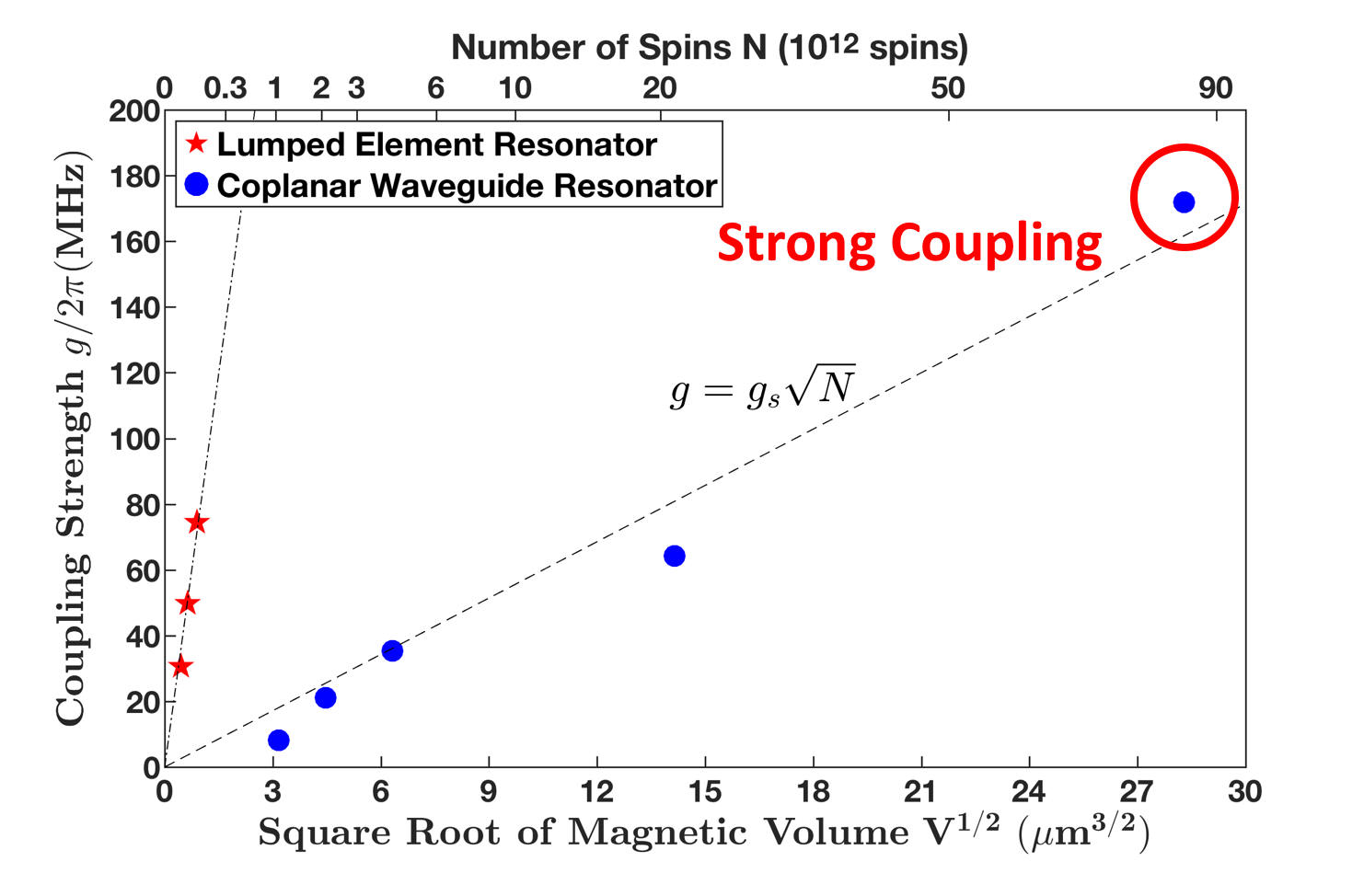}
    \caption{Magnon-photon coupling strength $g$ as a function of magnetic volume and spin number. The dashed lines represent fittings of the scaling rule $g=g_s\sqrt{N}$. The single spin coupling strength for two resonators in this work are determined to be $g_s^{\text{CPW}}/2\pi=18$ Hz and $g_s^{\text{LE}}/2\pi=263$ Hz in CPW resonators and lumped element resonators, respectively. }
    \label{fig3}
\end{figure}

Using $M_s$ and the magnetic volume of the Py stripe, we estimate the number of spins involved in the coupling to be $N=2.1 \times 10^{13}$. By fabricating devices with different length of Py stripes, we confirm the scaling of $g\propto \sqrt{N}$ and extract  $g_s^{\text{CPW}}/2\pi = 18$ Hz (Fig. \ref{fig3}). As the Py stripe is at the center of the 20 $\mu$m wide signal line, we can assume that the RF magnetic field is uniform throughout the Py volume and estimate $b_{\text{rf}}^{\text{CPW}} =\mu_0/2w^{\text{CPW}}$. Together with the designed impedance $Z_r^{\text{CPW}} \approx 50$ $\ohm$ of CPW resonators, we calculate the theoretical $g_s$ to be $g_{s,theory}^{\text{CPW}}/2\pi = 14$ Hz, which attains reasonable agreement with our experimental value.  In the device with 2000 $\mu$m long Py, coupling strength $g/2\pi = 171$ MHz is obtained which is larger than $\kappa_r$ and $\kappa_m$, and therefore falls into the strong coupling region. The corresponding cooperativity $g^2/\kappa_r\kappa_m=160$ is very high for this small magnetic volume.

Next, we adopt the strategy of impedance reduction to further enhance the coupling strength. Low-impedance lumped element LC resonator has been recently employed for paramagnetic electron spin resonance experiments \cite{eichler_electron_2017,bienfait_reaching_2016, bienfait_controlling_2016}, but the potential for reaching strong magnon-photon coupling remains largely unexplored. As is shown in Fig. \ref{fig2}a, the resonator consists of large inter-digitated capacitors in parallel with a small inductor, and is capacitively side-coupled to the signal line of a CPW. The measured transmission coefficient of this resonator is shown in Fig. \ref{fig2}b, where a minimum transmission shows up under the resonant condition due to its absorptive nature, in contrast to transmission peaks observed in the CPW resonator. Two resonant modes are observed in the transmission of the bare resonator, with resonant frequencies located at 5.42 and 9.19 GHz, separately. In order to understand the properties of the two modes, we carried out electromagnetic wave simulations (Sonnet) and found that the lower frequency mode corresponds to the case with a high current density passing through the central inductive wire (see simulation results in Fig. \ref{fig2}b). We estimate the capacitance $C$ of this mode analytically \cite{igreja_analytical_2004} to be 1.91 pF and obtain the corresponding inductance $L=0.45$ nH using the measured $\omega_r$. The characteristic impedance of this LC circuit is calculated to be $Z_r^{\text{LE}} = \sqrt{L/C} \approx15.3$ $\ohm$, much smaller than the value of CPW resonators. Moreover, the inductor width $w^{\text{LE}}$ is designed to be only 4 $\mu$m to further increase magnetic field intensity. Fig. \ref{fig2}c shows the transmission of a resonator that is coupled with a 40$\mu$m$\times$2$\mu$m$\times$10nm Py wire, as a function of frequency and applied magnetic field. Fitting the resonant frequencies evolution using Eq.\ \ref{twomodes}, we extract the coupling strength $g/2\pi=74.5$ MHz, the saturation magnetization $\mu_0 M_s=1.1$ T, and the resonator frequency $\omega_r^{\text{LE}}/2\pi = 5.253$ GHz. The relatively smaller $M_s$ value compared with that of the previous CPW resonator sample comes from the thinner Py film thickness (10 nm vs 50 nm) and potential magnetically dead interfacial layer \cite{ounadjela_thickness_1988}. With decay rates of the resonator mode $\kappa_r/2\pi=1.05$ MHz and magnon mode $\kappa_m/2\pi=122$ MHz, we calculate the cooperativity $g^2/\kappa_r\kappa_m = 43.3$, which is fairly large considering the very small number of spins ($N=7.3\times 10^{10}$). By fabricating devices with different length of Py stripes, we extract  $g_s^{\text{LE}}/2\pi = 263$ Hz (Fig. \ref{fig3}), which is an order of magnitude larger than the value with CPW resonator. The $g_s^{\text{LE}}/2\pi$ value obtained in our experiment is larger than the one calculated using our model $g_{s,theory}^{\text{LE}}/2\pi=141$ Hz, which can be attributed to the enhancement of magnetic field at the edge of the inductor wire due to the field's nonuniform distribution \cite{eichler_electron_2017, bienfait_controlling_2016}. The reasonable agreement between theoretical and experimental values in both CPW resonators and low-impedance lumped element resonators shows the usefulness of the formula $g=g_s\sqrt{N}=g_{e}\mu_B b_{\text{rf}} \omega_r \sqrt{N/8\hbar Z_r} $ obtained from our quantum mechanical model to predict magnon-photon coupling strength. 

In summary, we have demonstrated high cooperativity microwave photon-magnon coupling between a resonator mode in planar superconducting resonators and a Kittel mode in Py nanomagnets. With enhanced $g_s$, the number of spins $N$ involved for reaching strong coupling is 3 orders of magnitude lower than previous experiments. In our experiment, a ferromagnetic metal with relatively high damping coefficient (Py) is employed. By simply replacing magnetic metals with insulator thin films with ultralow damping such as YIG (Q$>$1000) \cite{chang_nanometer-thick_2014}, we expect strong magnon-photon coupling to be realized with as few as $10^7$ spins using our current design. On the other hand, our studies show that the coupling strength obtained from the analytical model provides relatively precise estimate on the experimental values, which can be used as guidelines for further scaling down the magnonic system volume. For example, a lumped element resonator made by nanofabrication technique \cite{haikka_proposal_2017} with inductor width of $\sim$100 nm can further enhance $g_s$ by a factor of 40 and reduce the number of spins for reaching strong coupling using YIG to $10^4$. Our system is on-chip, lithographically scalable \cite{kakuyanagi_observation_2016}, and Circuit QED compatible, which demonstrates high potential for integrated hybrid quantum systems harnessing magnon physics. The demonstration of the coupled systems with ferromagnetic metal provides the opportunities to investigate magnon-photon coupling in a wide range of spintronic devices, such as magnetic tunnel junctions. Moreover, the high coupling strength with nanomagnets opens up the possibility of electrical control of the hybrid system dynamics utilizing spintronic effects, such as spin torque \cite{kiselev_microwave_2003, ando_electric_2008, demidov_magnetic_2012, liu_magnetic_2012, duan_nanowire_2014} and voltage controlled magnetic anisotropy \cite{amiri_voltage-controlled_2012}.

This work is supported by National Science Foundation under award ECCS-1653553 and AFOSR under award FA9550-19-1-0048.

\bibliography{MagnonPhotonCoupling_arxiv_030419}

\newpage
\onecolumngrid
\section*{Supplemental Material}

\title{Supplement to ``Strong Coupling between Microwave Photons and Nanomagnet Magnons"}
\author{Justin T. Hou}
\affiliation{Department of Electrical Engineering and Computer Science, Massachusetts Institute of Technology, Cambridge, MA 02139, USA}
\author{Luqiao Liu}
\affiliation{Department of Electrical Engineering and Computer Science, Massachusetts Institute of Technology, Cambridge, MA 02139, USA}
\date{{\small \today}}
\maketitle
\section{Magnon-Photon Coupling Hamiltonian and Semi-classical Equations of Motion}

    We consider the quantum mechanical model of a quantum macrospin coupled to an LC resonator through oscillating magnetic field generated by the inductor. The macrospin model is justified as we observe the uniform ferromagnetic resonance mode, Kittel mode, in the experiment. The LC circuit model has been widely adopted to describe the photon modes of superconducting resonators in the context of Circuit QED [S1]. The Hamiltonian is composed of three parts: LC resonator $\hat{H}_r=L\hat{I}^2/2+C\hat{V}^2/2$, magnon $\hat{H}_m=g_{e}\mu_B/\hbar \mathbf{\hat{S}}\cdot \mathbf{B_{eff}}$, and interaction $\hat{H}_{int}=g_{e}\mu_B/\hbar \mathbf{\hat{S}}\cdot \mathbf{\hat{B}_{rf}}=g_{e}\mu_B/\hbar \mathbf{\hat{S}}\cdot \mathbf{b_{rf}} \hat{I}$, where $\mathbf{b_{rf}}$ is the magnetic field experienced by the macrospin per unit inductor current. The effective field of the magnon $\mathbf{B_{eff}}$ generally contains contributions from external field $\mathbf{B_{ext}}$ as well as magnetic anisotropy, which we neglect for simplicity. The macrospin operators satisfy the commutation relation $[\hat{S}_{i},\hat{S}_j]=i\hbar\epsilon_{ijk}\hat{S}_k$, where $\epsilon_{ijk}$ is the Levi-Civita symbol. To match our experimental setup, we adopt $\mathbf{B_{ext}}=-B_0 \mathbf{\hat{z}}$ and $\mathbf{b_{rf}}=b_{\text{rf}} \mathbf{\hat{x}}$. The total Hamiltonian can thus be written as :
\begin{align}
\begin{split}
\hat{H}=&\hat{H}_r+\hat{H}_m+\hat{H}_{int} \\
=&\frac{1}{2}L\hat{I}^2+\frac{1}{2}C\hat{V}^2+\frac{g_{e}\mu_B}{\hbar}\mathbf{\hat{S}}\cdot \mathbf{B_{ext}}+\frac{g_{e}\mu_B}{\hbar}\mathbf{\hat{S}}\cdot \mathbf{\hat{B}_{rf}}\\
=&\frac{1}{2}L\hat{I}^2+\frac{1}{2}C\hat{V}^2-\frac{g_{e}\mu_B}{\hbar} B_0\hat{S}_z+\frac{g_{e}\mu_Bb_{\text{rf}}}{\hbar} \hat{S}_x \hat{I} 
\end{split}
\end{align}

Note that without the interaction Hamiltonian $\hat{H}_{int}$, the ground state of the macrospin is with $\braket{\mathbf{\hat{S}}}=\hbar S\hat{\mathbf{z}}$ where the macrospin with total spin $S$ is aligned in positive z-direction by the external field. To proceed, we express the macrospin operator $\mathbf{\hat{S}}=\frac{1}{2}(\hat{S}_{+}+\hat{S}_{-}) \mathbf{\hat{x}}+\frac{1}{2i}(\hat{S}_{+}-\hat{S}_{-}) \mathbf{\hat{y}}+\hat{S}_z \mathbf{\hat{z}}$, with $\hat{S}_{+}$ ($\hat{S}_{-}$) raising (lowering) the $z$-component of the macrospin. We also express operators $\hat{V}$ and $\hat{I}$ in terms of creation (annihilation) operators of photon modes in the resonator $\hat{a}^{\dagger}_r $ ($\hat{a}_r$): $\hat{V}=i\omega_r\sqrt{\hbar Z_r/2} (\hat{a}^{\dagger}_r-\hat{a}_r)$ and $\hat{I}=\omega_r\sqrt{\hbar/2Z_r} (\hat{a}^{\dagger}_r+\hat{a}_r)$, where $Z_r=\sqrt{L/C}$ and $\omega_r=1/\sqrt{LC}$ are the characteristic impedance and the resonant frequency of the LC resonator, respectively. Note that in this LC circuit quantization convention, the semiclassical equations of motion without the macrospin are  $d\braket{\hat{V}}/dt=\braket{[\hat{V},\hat{H}_r]}/i\hbar=-\braket{\hat{I}}/C$ and $d\braket{\hat{I}}/dt=\braket{[\hat{I},\hat{H}_r]}/i\hbar=\braket{\hat{V}}/L$ obtained by Ehrenfest Theorem ($d\braket{\hat{O}}/dt=\braket{[\hat{O},\hat{H}]}/i\hbar$ for general time independent operator $\hat{O}$), which resembles dynamics of a classical LC circuit. Moreover, the equations of motion of the macrospin without the LC resonator can be obtained as $d\braket{\hat{S}_x}/dt=\braket{[\hat{S}_x,\hat{H}_m]}/i\hbar=g_e\mu_BB_0\braket{\hat{S}_y}/\hbar$ and $d\braket{\hat{S}_y}/dt=\braket{[\hat{S}_y,\hat{H}_m]}/i\hbar=-g_e\mu_BB_0\braket{\hat{S}_x}/\hbar$, which represents circular precession of the magnon mode. With the above expressions of operators and adopting rotating wave approximation, the Hamiltonian can now be written as:
\begin{align}
\begin{split}
\hat{H}=&\hbar \omega_r (\hat{a}^{\dagger}_r\hat{a}_r+\frac{1}{2})-\frac{g_{e}\mu_B}{\hbar} B_0\hat{S}_z+\frac{g_{e}\mu_B b_{\text{rf}}\omega_r}{\hbar} \sqrt{\frac{\hbar}{8Z_r}}(\hat{S}_+\hat{a}^{\dagger}_r+\hat{S}_-\hat{a}_r)\\
=&\hbar \omega_r (\hat{a}^{\dagger}_r\hat{a}_r+\frac{1}{2})-\omega_m(B_0)\hat{S}_z+g_s(\hat{S}_+\hat{a}^{\dagger}_r+\hat{S}_-\hat{a}_r)
\end{split}
\end{align}
where $\omega_m(B_0)=g_{e}\mu_BB_0/\hbar$ and $g_s=g_{e}\mu_B b_{\text{rf}} \omega_r/\sqrt{8\hbar Z_r}$ are the magnon frequency (a function of applied field $B_0$) and single spin-photon coupling strength which will become clear shortly, respectively. We can now compute the time evolution of expectation values of resonator and macrospin operators using Ehrenfest Theorem:

\begin{align}
\begin{split}
\frac{d\braket{\hat{a}_r}}{dt}&=-i\omega_r \braket{\hat{a}_r}-i\frac{g_s}{\hbar}\braket{\hat{S}_+} \\
\frac{d\braket{\hat{S}_+}}{dt}&=-i\omega_m(B_0) \braket{\hat{S}_+}-i 2g_s\braket{\hat{a}_r \hat{S}_z} \\
\frac{d\braket{\hat{a}^{\dagger}_r}}{dt}&=i\omega_r \braket{\hat{a}^{\dagger}_r}+i\frac{g_s}{\hbar}\braket{\hat{S}_-} \\
\frac{d\braket{\hat{S}_-}}{dt}&=i\omega_m(B_0) \braket{\hat{S}_-}+i 2g_s\braket{\hat{a}^{\dagger}_r \hat{S}_z}
\end{split}
\end{align}

Here we assume that the amplitude of magnetic oscillation is small such that the state is approximately an eigenstate of $\hat{S}_z$ with eigenvalue $\hbar S=\frac{\hbar N}{2}$, where $N$ is the total number of spins mentioned in the main text. In this assumption, the upper two equations are decoupled from the lower two equations. Moreover, the algebraic structure is similar to coupled harmonic oscillators. We assume harmonic time dependence $\braket{\hat{a}_r}=a_0e^{i\omega t}$ and $\braket{\hat{S}_+}=S_0e^{i\omega t}$. The upper two equations indicate the eigenvalue problem:
\begin{equation}
 \begin{pmatrix} \omega-\omega_r & g_s/\hbar \\ Ng_s\hbar & \omega-\omega_m(B_0) \end{pmatrix} \binom{a_0}{S_0} =0
\end{equation}
The eigenvalue equation gives two resonant frequencies: $\omega_{\pm}=\omega_r+\frac{\Delta}{2}\pm\sqrt{\Delta^2+4g^2}$, where $\Delta=\omega_m(B_0)-\omega_r$ is the detuning and $g=g_s\sqrt{N}$ is the total magnon-photon coupling strength. The coupled modes dispersion describes the characteristic anticrossing observed in experiment. Here we note the scaling of coupling strength with number of spins $g=g_s\sqrt{N}$, with the single spin coupling strength $g_s=g_{e}\mu_B b_{\text{rf}} \omega_r/\sqrt{8\hbar Z_r}$ derived in a quantum mechanical manner. The semiclassical model expresses classical physical quantities (such as $I(t)$ and $S_x(t)$) in terms of quantum expectation values of operators (such as $\braket{\hat{I}}$ and $\braket{\hat{S}_x}$), whose time evolutions are related to those of $\braket{\hat{a}_r}$ and $\braket{\hat{S}_+}$, together with $\braket{\hat{a}^{\dagger}_r}=\braket{\hat{a}_r}^{*}$ and $\braket{\hat{S}_-}=\braket{\hat{S}_+}^{*}$. The constants $a_{0\pm}$ and $S_{0\pm}$ can be determined by initial conditions of the problem in consideration. This can be adopted to describe "Rabi-like" oscillations in magnon-photon coupled systems in classical region. [S2]

We can also rewrite the Hamiltonian using Holstein-Primakoff transformation which expresses the macrospin operators as magnon excitation boson operators $\hat{a}^{\dagger}_m$ and $\hat{a}_m$:
\begin{align}
\begin{split}
\hat{S}_-=&\hbar \hat{a}^{\dagger}_m\sqrt{2S-\hat{a}^{\dagger}_m\hat{a}_m}\\
\hat{S}_+=&\hbar \sqrt{2S-\hat{a}^{\dagger}_m\hat{a}_m}\hat{a}_m \\
\hat{S}_z=&\hbar S-\hbar \hat{a}^{\dagger}_m\hat{a}_m 
\end{split}
\end{align}
Again, we assume the magnon excitation is small $(\braket{\hat{a}^{\dagger}_m\hat{a}_m}/S\ll 1)$ and neglect the term $\hat{a}^{\dagger}_m\hat{a}_m$ in the transformation for $\hat{S}_{-}$ and $\hat{S}_{+}$. The total spin is $S=N/2$. We drop the constant energy terms and get the Hamiltonian:
\begin{equation}
H=\hbar \omega_r \hat{a}^{\dagger}_r\hat{a}_r+\hbar \omega_m(B_0) \hat{a}^{\dagger}_m\hat{a}_m+\hbar g_s\sqrt{N}(\hat{a}_m\hat{a}^{\dagger}_r+\hat{a}^{\dagger}_m\hat{a}_r) 
\end{equation}
The time evolution of the operators can be computed:
\begin{align}
\begin{split}
\frac{d\braket{\hat{a}_r}}{dt}&=-i\omega_r \braket{\hat{a}_r}-ig_s\sqrt{N}\braket{\hat{a}_m} \\
\frac{d\braket{\hat{a}_m}}{dt}&=-i\omega_m(B_0) \braket{\hat{a}_m}-i g_s\sqrt{N}\braket{\hat{a}_r} \\
\frac{d\braket{\hat{a}^{\dagger}_r}}{dt}&=i\omega_r\braket{\hat{a}^{\dagger}_r}+ig_s\sqrt{N}\braket{\hat{a}^{\dagger}_m} \\
\frac{d\braket{\hat{a}^{\dagger}_m}}{dt}&=i\omega_m(B_0)\braket{\hat{a}^{\dagger}_m}+ig_s\sqrt{N}\braket{\hat{a}^{\dagger}_r} 
\end{split}
\end{align}
The upper two equations are already decoupled from the lower two, due to the assumption of small magnon excitation. We assume harmonic time dependence $\braket{\hat{a}_r}=a_{r0}e^{i\omega t}$ and $\braket{\hat{a}_m}=a_{m0}e^{i\omega t}$. We get eigenvalue equation:
\begin{equation}
 \begin{pmatrix} \omega-\omega_r & g_s\sqrt{N} \\ g_s\sqrt{N} & \omega-\omega_m(B_0) \end{pmatrix} \binom{a_{r0}}{a_{m0}} =0
\end{equation}
We again obtain $\omega_{\pm}=\omega_r+\frac{\Delta}{2}\pm\sqrt{\Delta^2+4g^2}$, where $g=g_s\sqrt{N}$ is the total magnon-photon coupling strength.

\vskip 0.1in
\vskip 0.1in
[S1] M. H. Devoret, Quantum Fluctuations in Electrical Circuits. Elsevier, page 372, 1997. 
\vskip 0.1in
[S2] X. Zhang, C.-L. Zou, L. Jiang, and H. X. Tang, Physical Review Letters $\textbf{113}$, 156401 (2014).

\end{document}